\documentclass[apj]{emulateapj}
\usepackage[colorlinks,linkcolor={blue},citecolor={blue},urlcolor={red}]{hyperref}
\usepackage{mathtools}
\usepackage{epsfig,graphicx,natbib,amsmath,amsfonts,amssymb}
\usepackage{color}
\usepackage{multirow}
\usepackage{booktabs}
\usepackage{amssymb}
\usepackage{booktabs} 
\usepackage{mathrsfs}  
\usepackage{longtable}
\usepackage{array} 
\usepackage[dvipsnames]{xcolor}  
\usepackage{lineno}

\newcommand{\msolar}{${\rm M}_\odot$}

\newcommand{\myemail}{\email{enci.wang@phys.ethz.ch}}

\shorttitle{The metallicity profiles of exponential star-forming disk}
\shortauthors{Wang \& Lilly}

\graphicspath{{fig/}}

\begin{document}
\title {The gas-phase metallicity profiles of star-forming galaxies in the modified accretion disk framework}

\author {Enci Wang\altaffilmark{1},
Simon J. Lilly\altaffilmark{1}
} \myemail

\altaffiltext{1}{Department of Physics, ETH Zurich, Wolfgang-Pauli-Strasse 27, CH-8093 Zurich, Switzerland}

\begin{abstract}  

Simulations indicate that the inflow of gas of star-forming galaxies is almost co-planar and co-rotating with the gas disk, and that the outflow of gas driven by stellar winds and/or supernova explosions is preferentially perpendicular to the disk.  This indicates that the galactic gas disk can be treated as a {\it modified} accretion disk.  In this work, we focus on the metal enhancement in galactic disks in this scenario of gas accretion.  
Assuming that the star formation rate surface density ($\Sigma_{\rm SFR}$) is of exponential form, we obtain the analytic solution of gas-phase metallicity with only three free parameters: the scalelength of $\Sigma_{\rm SFR}$ ($h_{\rm R}$), the metallicity of the inflowing gas and the mass-loading factor defined as the wind-driven outflow rate surface density per $\Sigma_{\rm SFR}$.  
According to this simple model, the negative gradient of gas-phase metallicity is a natural consequence of the radial inflow of cold gas which is continuously enriched by in-situ star formation as it moves towards the disk center.   
We fit the model to the observed metallicity profiles for six nearby galaxies chosen to have well-measured metallicity profiles extending to very large radii. 
Our model can well characterize the overall features of the observed metallicity profiles.  The observed profiles usually show a floor at the outer regions of the disk, corresponding to the metallicity of inflow gas.  Furthermore, we find the $h_{\rm R}$ of $\Sigma_{\rm SFR}$ inferred from these fits agree well with independent estimates from $\Sigma_{\rm SFR}$ profiles, supporting the basic model.  

\end{abstract}

\section{Introduction}
\label{sec:introduction}


Cold gas accretion is essential to sustain the star formation and size growth of disk galaxies during their evolution \citep[e.g.][]{Binney-00, Keres-05, Dekel-06, Sancisi-08, Silk-12, Conselice-13, Sanchez-Almeida-14, Trapp-21}. Two modes of accretion have been proposed: the smooth accretion from the circumgalactic medium (CGM) and mergers with dwarf companions \citep[e.g.][]{Lacey-94, Murali-02, Keres-09, Bouche-10, LHuillier-12, Sanchez-Almeida-14, Rodriguez-Gomez-15}. Multi-zoom cosmological simulations suggest that galaxies assemble their mass mainly via smooth accretion, and that mergers are important only for the most massive high-redshift galaxies \citep{Murali-02, Keres-09, LHuillier-12}. Since the smooth accretion dominates the gas inflow, how the gas flows onto the galactic disk is therefore the key to understanding the formation and evolution of disks and their chemical enrichment.  

Although there is not yet a lot of direct observational evidence for smooth accretion, hydrodynamical simulations can be a powerful guide.  Gas can evidently accrete in a ``cold-mode'' along filamentary streams without being shock-heated to the virial temperature in the outer halo \citep{Keres-05, Dekel-06, Ocvirk-08, Brooks-09, vandeVoort-11, Stern-20}.  This kind of accretion dominates for low mass galaxies at high-redshift. The accretion of cold gas from the cooling of hot halo gas may be more important in more massive disk galaxies \citep{Keres-05, Ocvirk-08, Nelson-13, Stern-20}. More importantly, there is growing evidence in the simulations that the inflowing gas is almost co-planar and co-rotating with the gas disk, regardless of its thermal history \citep[e.g.][]{Keres-05, Danovich-15, Stewart-17, Stern-20, Peroux-20, Trapp-21}, especially at low-redshift, where the turbulent motion of gas is not likely to be significant.  In contrast, the outflowing gas that is driven by stellar winds and/or supernova (SN) explosions leaves the disk following the path of least resistance, i.e. preferentially along the direction that is perpendicular to the disk \citep[e.g.][]{Peroux-20, Trapp-21}. Such outflow may further inhibit the infall of material from regions above and below the disk plane.  

Observationally, there is indirect evidence to support the above picture of gas accretion. 
By mapping the circumgalactic medium (CGM) with the Mg II absorption, disk galaxies are found to have the co-planar gaseous structure in the CGM plus a strongly bipolar outflow aligned along the disk rotation axis \citep[also see][]{Bordoloi-11, Bouche-12, Kacprzak-12, Schroetter-19}. The co-planar gaseous structure in CGM is found to co-rotate with the galactic disk \citep{Diamond-Stanic-16,  Bielby-17, Peroux-17}.

The nature of this gas accretion suggests that the gas disk of galaxies can be treated as an {\it modified} accretion disk. The ``modified'' here emphasizes the differences from the classical accretion disk of the black holes in that the inflow rate radially decreases towards the galactic center as gas is consumed by star-formation or removed by associated outflows.  In the first paper in this series 
\citep[][hereafter Paper I]{Wang-22}, we have investigated the possible mechanisms for the radial transport of gas in the framework of a modified accretion disk. We found that magneto-rotational instability (MRI) provides an effective source of viscosity for the transportation of mass and angular momentum within the gas disk, and is likely responsible for the formation and evolution of the observed exponential profiles of the star-forming (SF) disks. 

In the present work, we focus on the chemical enrichment of the gas disk in this modified accretion disk framework.  Our goal is to determine whether or not the modified accretion disk model can reasonably reproduce the observed metallicity gradients in galaxies.  This not only may give insight into the origin of metallicity profiles in galaxies, but also provides an independent test for the validity of the modified accretion disk model that was  introduced in Paper I.  We stress however that the present work is quite independent of the source of viscosity in that model and should hold for any model of disk galaxies in which the dominant gas flow is radial inflow within the plane of the disk.

The implications on the chemical enrichment of a dominant radial inflow of accreted gas within a disk have not attracted a lot of attention in the literature. For instance, \cite{Belfiore-19} modeled the radial profiles of gas-phase metallicity for low-redshift star-forming galaxies, in a gas-regulator framework in which the cold gas content of a galaxy is regulated by the interplay between inflow, star formation and outflow \citep[also see][]{Lilly-13, Wang-19, Wang-21}. However, in their work, the radial inflow of cold gas along the disk is fully ignored \citep[e.g.][]{Chiappini-01, Schonrich-17, Lian-18, Belfiore-19}, i.e. all the annuli within the disk were assumed to evolve independently.  In this case, the radial gradient of the gas-phase metallicity in the disk is determined by the radial dependence of the mass-loading factor and the metallicity of the inflowing gas at steady-state, as illustrated in \cite{Wang-21}. The SFE may also play a role if the system is not in steady-state, because the SFE determines how quickly the system responds to changes in the inflow rate \citep{Peng-14, Wang-19}. However, the directions of cold gas inflow and outflow would definitely have an effect on the metal-enrichment at different galactic radii, i.e. on the radial gradient of the gas-phase metallicity \citep[e.g.][]{Mayor-81, Lacey-85, Sommer-Larsen-89,  Goetz-92, Thon-98, Portinari-00, Spitoni-11, Cavichia-14, Kubryk-15, Pezzulli-16, Sharda-21}. 

The chemical enhancement of viscous galactic disks has been investigated in the literature \citep[e.g.][]{Lacey-85, Lin-87, Sommer-Larsen-89, Sommer-Larsen-90, Firmani-96, Thon-98, Spitoni-11, Sharda-21}. However, these works still included strong assumptions about the gas inflow from ex-planar gas, and the metallicity of the inflow, at different galactic radii, in order to reproduce the observed SFR surface density profiles and/or the metallicity profile. Since the inflow of gas is likely to be primarily co-planar, as indicated by recent simulations \citep[e.g.][]{Peroux-20, Trapp-21}, we consider in this work a simplified model in which the co-planar radial inflow of gas dominates the gas accretion and the outflow driven by stellar winds and supernova feedback is ex-planar.  In other words, the star formation and the size growth of disk galaxies are sustained by the radial migration of co-planar gas moving inwards within the disk from the CGM. 

In addition, we will generally assume as an input that the profile of the star formation rate (SFR) surface density, $\Sigma_{gas}(r)$ is exponential, as typically found in SF galaxies \citep[e.g.][Paper I]{Wyder-09, Gonzalez-Delgado-16, Casasola-17, Wang-19}. This avoids introducing more free parameters, and enables us to solve the radial inflow and gas-phase metallicity profiles analytically (see more details in Section \ref{sec:2} and \ref{sec:3}).  We will not need to assume anything about the radial distribution of gas.

The paper is organized as follows. In Section \ref{sec:2}, we present the detailed model assumptions and the basic equations. In Section \ref{sec:3} and \ref{sec:4}, we present the solutions for gas-phase metallicity profile that are obtained with the assumed exponential SF disk. In Section \ref{sec:5}, we apply the model prediction to the observed gas-phase metallicity profiles, which gives a new explanation for the flattening of metallicity at the outer regions of galactic disks.  We discuss the role of star-formation efficiency and the validation of our model in Section \ref{sec:6}.  We summarise this work in Section \ref{sec:7}.   Throughout the paper, we use the term metallicity to refer to the gas-phase metallicity. 
We adopt a flat cold dark matter cosmology model with $\Omega_m=0.27$, $\Omega_\Lambda=0.73$ and $H_0 = 70$ km s$^{-1}$Mpc$^{-1}$
when computing distance-dependent parameters.

\section{The radial metal enrichment model in the modified accretion disk frame}  \label{sec:2}

\subsection{The modified accretion disk model} \label{sec:2.1}

The basic modified accretion disk model was presented in Paper I, and here we only briefly summarize the key points of this model. The reader is referred to that paper for more details.

First, co-planar gas inflow is assumed to dominate the smooth accretion of gas \citep[e.g.][]{Murali-02, Keres-09, LHuillier-12, Peroux-20, Trapp-21}. This radial co-planar flow is the main source of fuel for star formation and for any associated outflows.   

Second, it is assumed for simplicity that the star-formation driven outflow and any ex-planar {\it inflow} are both proportional to the instantaneous local $\Sigma_{\rm SFR}$.  This means that they may be combined into a single effective outflow, given by the instantaneous $\Sigma_{\rm SFR}$ multiplied by a mass-loading factor $\lambda$. 
In terms of chemical enrichment, since the ex-planar inflow is likely to be material that was originally from the gas disk \citep{Fraternali-15}, we also assume, again for simplicity, that the metallicity of this ex-planar inflow is the same as the metallicity of the gas disk at the corresponding radius.
The effective $\lambda$ is also assumed to be the same at all galactocentric radii, but that it can evolve with time and may vary from galaxy to galaxy.  
In effect, we are ignoring any mixing of exoplanar gas in this simple model.

Third, we assume that the galactic gas disk behaves as a gas-regulator system \citep[e.g.][]{Sommer-Larsen-89, Thon-98, Bouche-10, Schaye-10, Lilly-13, Wang-18, Wang-21},
in which the local SFR surface density $\Sigma_{\rm SFR}$ is instantaneously determined by the local $\Sigma_{\rm gas}$ which is itself the result of the interplay between continuous gas inflow, star formation, and any associated outflows. 
The star-formation law defines at each point a star-formation efficiency, SFE, which is simply ${\rm SFE} = \Sigma_{\rm SFR} / \Sigma_{\rm gas}$ at that location. 
The fraction of mass formed in new stars that is subsequently returned to the interstellar medium through winds and supernova explosion is denoted as $R$. We adopt the instantaneous return assumption and take $R =$ 0.4 with \cite{Chabrier-03} initial mass function (IMF) from stellar population models \citep[][]{Bruzual-03, Vincenzo-16}. 

In Paper I, we also assumed that the gas disk is everywhere rotationally supported, but the present work is free from this assumption.   As in Paper I, we do assume here that the system is rotationally symmetric, and consider only the azimuthally-averaged radial variations of all quantities.

\subsection{The treatment of metal enrichment} \label{sec:2.2}

Compared to the previous chemical evolution models of galactic disks \citep{Lacey-85, Lin-87, Sommer-Larsen-89, Sommer-Larsen-90,  Thon-98, Kubryk-15, Pezzulli-16, Lian-18, Belfiore-19},  there are two main differences in the present work. First,  since the co-planar radial inflow dominates the gas accretion, 
and since we assume that ex-planar inflow and outflow have the same metallicity,
we can combine the outflow and inflow from ex-planar gas into a single ``effective" outflow \citep{Fraternali-15}. This avoids assuming an arbitrary ex-planar inflow of gas. 
Second, the $\Sigma_{\rm SFR}(r)$ profile is assumed to have a (precise) exponential form.  The exponential form is motivated by  observations \citep[][Paper I]{Gonzalez-Delgado-16, Casasola-17, Wang-19}.  
These two assumptions largely reduce the degrees of freedom of the model, and enable us to present analytic solutions for the radial metallicity profiles of the gas disks (see Section \ref{sec:3}). 

Since the lifetime of the massive stars that collapse into core-collapse supernova (a few Myrs) is negligible to the other timescales of relevance, we make the instantaneous metal-enrichment and the instantaneous mixing approximations. This means we effectively neglect the ejection of metals by Type-Ia supernova \citep{Mannucci-NG}.  This is a common assumption in many chemical evolution models. After a single episode of star formation, more than 80\% (or 90\%) of Oxygen (with respect to total released Oxygen) are released within 10 Myr (or 20 Myr) by the core-collapse supernova \citep[see figure 1 in][]{Maiolino-19}, while the contribution of Type-Ia supernova is negligible. This makes the assumption of instantaneous metal-enrichment reasonable for Oxygen but not for some other elements, like Carbon, Nitrogen and iron, because the timescale for production of these elements are much longer (on $\sim$Gyr timescales), and because the contribution to the production of these elements by Type-Ia supernova can be as large as a few tens of percent \citep{Maiolino-19}.  Observationally, the abundance of Oxygen is widely used to trace the gas-phase metallicity, which enables the direct comparison with our model prediction.   

As noted above, we assume that ex-planar flows, both outflow and inflow have the same metallicity as the disk at the same radius.   While the assumption about the outflow is not unreasonable (neglecting the possibility that the supernova-driven outflow is preferentially enriched) assigning the same metallicity to the inflow is tantamount to assuming that there is little mixing of gas in the halo of the galaxy.  Overall this would not be a concern if the gas exchange between ex-planar gas and the disk is insignificant, i.e. if the explanar inflow and outflow rate are both small with respect to the SFR.  However, this gas exchange may be significant for low mass galaxies because of their large mass-loading factors \citep{Lilly-13, Muratov-15}. This means that the significant local gas recycling out of the plane of the disk may smooth out the radial profile of gas-phase metallicity for lower mass galaxies.

Likewise, we will also not, at least initially, consider radial mixing of gas within the disk, i.e. radial diffusion of metals.  The effect of radial diffusion will be explicitly considered in Section \ref{sec:3.3} below.

The model is constructed in one-dimension, i.e. we integrate over the vertical structure of the disk. In addition, we neglect azimuthal structures like spiral arms. 
We note that the co-planar net inflow rate $\Phi$, the SFR surface density $\Sigma_{\rm SFR}$, the mass surface density of cold gas $\Sigma_{\rm gas}$, and the gas metallicity $Z$, are functions of both radius and time. 

\subsection{Basic equations} \label{sec:2.3}

The basic continuity equation of gas mass at a given radius can be written as \citep[also see][ Paper I]{Lacey-85, Sommer-Larsen-89}: 
\begin{equation} \label{eq:1}
    \frac{\partial \Sigma_{\rm gas}}{\partial t} = \frac{\partial \Phi}{2\pi r \partial r} - (1-R+\lambda)\cdot \Sigma_{\rm SFR}.
\end{equation}
In the right-hand side of Equation \ref{eq:1}, the first term is the change rate of $\Sigma_{\rm gas}$ due to the co-planar radial inflow, and the second term is the change rate of  $\Sigma_{\rm gas}$ due to star formation and outflow. We emphasize that the $\Phi(r)$ is the {\it net} radial inflow rate. 

We denote the yield, i.e. the mass of metals returned to the interstellar medium per unit mass of instantaneously formed stars, as $y$. In a similar way as above, the basic continuity equation for the mass of metals ($Z\Sigma_{\rm gas}$) can be written as: 
\begin{equation} \label{eq:2}
    \frac{\partial (Z\cdot \Sigma_{\rm gas})}{\partial t} = \frac{\partial(\Phi\cdot Z)}{2\pi r \partial r} - Z\cdot (1-R+\lambda)\Sigma_{\rm SFR} + y\cdot \Sigma_{\rm SFR}. 
\end{equation}
In the right-hand of Equation \ref{eq:2}, the first term comes from co-planar inflow, the second term comes from the star formation and outflow, and the third term comes from the metal production in star formation. Combining Equation \ref{eq:1}, Equation \ref{eq:2} can be simplified as 
\begin{equation} \label{eq:3}
    \Sigma_{\rm gas} \cdot \frac{\partial Z}{\partial t} = \Phi \cdot \frac{\partial Z}{2\pi r \partial r} + y\cdot \Sigma_{\rm SFR}
\end{equation}
In principle, there should be an additional term representing standard diffusion in the right-hand of Equation \ref{eq:3}  \citep[e.g.][]{Sommer-Larsen-90, Sharda-21}. We ignore the diffusion term in our fiducial model, and explore the effect of diffusion in Section \ref{sec:3.3}. We note that there is no diffusion term in Equation \ref{eq:1}, because the $\Phi$ we considered is the {\it net} inflow rate. 

The observed gas-phase metallicity is usually the Oxygen abundance, defined as the number ratio of Oxygen to Hydrogen, rather than the mass ratio. We argue that in Equations \ref{eq:2} or \ref{eq:3}, the mass of metals can be replaced by the number of Oxygen, if the yield is defined in the terms of the number fraction of Oxygen.

The Equations \ref{eq:1} and \ref{eq:2} are the two main equations of this work, and are only based on the continuity of the mass of gas and metals. Given the basic assumptions of our model (see Section \ref{sec:2.1}), these two equations are well established and independent of the underlying physical viscous processes that drive the gas inflow.  In next section, we will explore the solutions of Equation \ref{eq:1} and \ref{eq:2}, which give the radial dependence of the co-planar inflow rate and gas-phase metallicity. 

\section{The predicted steady-state metallicity profile} \label{sec:3}

Based on the equations in Section \ref{sec:2.3}, we first investigate the analytic solutions of the co-planar inflow rate and gas-phase metallicity in the steady state of the gas disk, i.e. 
\begin{equation} \label{eq:4}
    \frac{\partial \Sigma_{\rm gas}}{\partial t} =0, {\rm and} \ 
    \frac{\partial Z}{\partial t} = 0
\end{equation}
This means that in the steady-state, both the gas surface density $\Sigma_{\rm gas}(r)$ and the metallicity $Z(r)$ of the gas disk will not evolve with time \citep[][also see Paper I]{Bouche-10, Lilly-13, Wang-21}.  However, we know this may not be true for SF galaxies on cosmological timescales, since strong evolutions of SFR and gas-phase metallicity are seen from observations \citep{Croom-12, Bundy-15, Huang-19, Gillman-21}, especially at high redshift.  We will consider the case of cosmological evolution in our framework in Section \ref{sec:4} below.

\subsection{The general solutions at steady-state} \label{sec:3.0}

Inserting Equation \ref{eq:4} into Equation \ref{eq:1}, we can obtain the general solution for the steady-state inflow: 
\begin{equation} \label{eq:xx1}
\Phi(r) =  \int_0^r 2\pi r' \cdot (1-R+\lambda)\Sigma_{\rm SFR}(r') dr' + \Phi(0),
\end{equation}
or
\begin{equation} \label{eq:xx11}
\Phi(r) = \Phi(+\infty) - \int^{+\infty}_{r} 2\pi r' \cdot (1-R+\lambda)\Sigma_{\rm SFR}(r') dr'.
\end{equation}
These two equations are fully equivalent. The first one indicates that the inflow rate at radius $r$ equals to the rate of star formation and outflow within this radius plus the mass rate of any central sink. The second one indicates that the inflow rate at radius $r$ equals the inflow rate at infinity minus the rate of star formation and outflow between infinity and the radius $r$. 

Similarly, inserting Equation \ref{eq:4} into Equation \ref{eq:3}, the solution of gas-phase metallicity in the steady-state can be written as:
\begin{equation} \label{eq:xx2}
Z(r) = \int_{+\infty}^{r} -\frac{2\pi r' \cdot y\cdot \Sigma_{\rm SFR}(r')}{\Phi(r')} dr' + Z_0. 
\end{equation}
This equation directly shows that the metallicity at given radius $r$ in the disk equals the metallicity of original inflow gas $Z_0$ plus the metal-enrichment due to star formation from infinity to radius $r$ when moving toward the disk center.  The outer boundary of the disk is set to be infinity in Equations \ref{eq:xx11} and \ref{eq:xx2} for simplicity, but one could certainly set a physical boundary to the disk instead. 

As shown in Equations \ref{eq:xx1}, \ref{eq:xx11} and \ref{eq:xx2}, the detailed form of $\Sigma_{\rm SFR}(r)$ is needed to determine the solutions for both the inflow rate and the metallicity. 
Since the $\Sigma_{\rm SFR}$ is well characterized by the exponential function for most SF galaxies (see Paper I), we look in this work for the solutions for a purely exponential SF disk: 
\begin{equation} \label{eq:5}
 \Sigma_{\rm SFR} = \Sigma_{0}\cdot e^{-r/h_{\rm R}}, 
\end{equation}
where $\Sigma_0$ is the SFR surface density at the center of the disk, and $h_{\rm R}$ is the exponential scalelength of the $\Sigma_{\rm SFR}$ profile.  As discussed later, we do not need to specify the radial gas content $\Sigma_{\rm gas}(r)$ of the disk.

\subsection{The solution of radial inflow rate}

We obtained the analytic solution of $\Phi$ with exponential $\Sigma_{\rm SFR}$ in Paper I and discussed it in some detail. Therefore, here we only present the results, and refer the readers to Paper I for more details. 
Substituting Equation \ref{eq:5} into Equation \ref{eq:xx1}, we directly obtain the analytic solution of the inflow rate as: 
\begin{equation} \label{eq:6}
     \frac{\Phi(r)}{1-R+\lambda} = {\rm SFR} \cdot [1+\eta-(x+1)\cdot \exp(-x)], 
\end{equation}
where $x$ is a scaled radius defined as $x=r/h_{\rm R}$, the SFR refers to the integrated SFR of the whole gas disk (SFR = $2\pi \Sigma_0 h_{\rm R}^2$), and the factor $\eta$ accounts for any mass sink at the center of the disk.  This could, for example, represent black hole accretion and jet-driven outflow at galactic center. 
We assume that this central mass sink is a (small) factor $\eta$ of the overall accretion rate that is required by the integrated star formation and wind-driven outflows, i.e. $\Phi_{\rm BH} = \eta \cdot (1-R+\lambda)$SFR. 

Consistent with Paper I, we denote the central mass sink as being due to a central massive black hole, but there are surely other possibilities also.  The mass sink $\eta$ is likely to be of order or greater than $\sim0.001$, i.e. the typical ratio of black halo mass and stellar mass of galaxies.  As expected, the $\Phi(r)$ monotonically 
increases with radius, approaching the overall accretion rate at large radii (see section 2.3 in Paper I). 


\subsection{The solution for the gas-phase metallicity}

\begin{figure*}
  \begin{center}
    \epsfig{figure=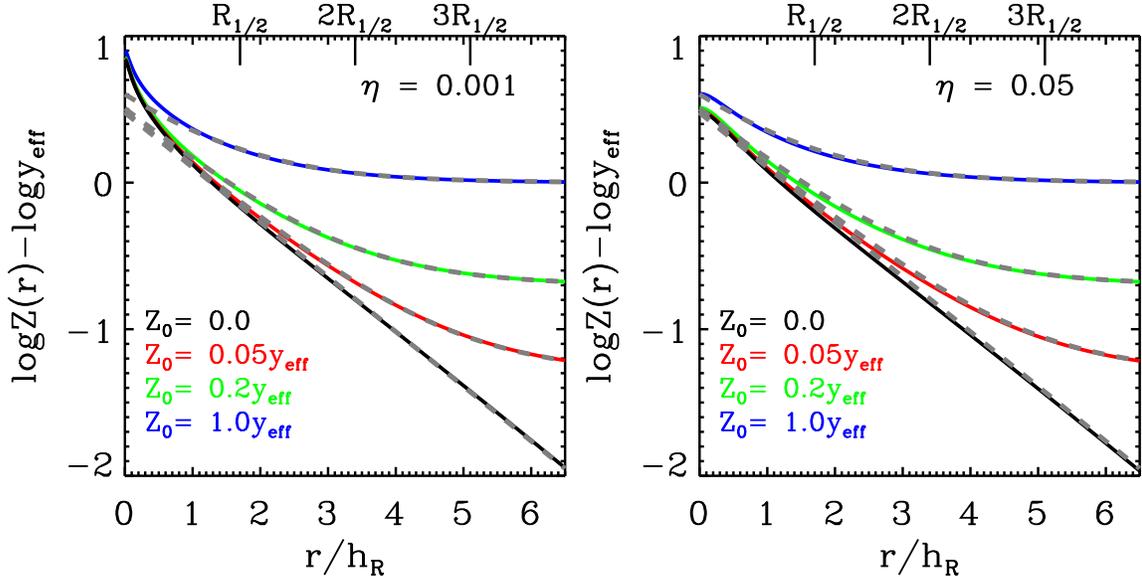,clip=true,width=0.9\textwidth}
    \end{center}
  \caption{The steady-state solution of the gas-phase metallicity profile in the modified accretion disk model, plotted as a function of radius normalized by the scale-length $h_{\rm R}$ of the exponential star-forming disk. We present the gas-phase metallicity scaled by the $y_{\rm eff}$, because the shape of $\log Z(r) -\log y_{\rm eff}$ profile does not depend on the settings of the yield $y$ and mass-loading factor $\lambda$.
  The solid curves are exact solutions at different $Z_0$, while the dashed gray curves are the approximated solution (see Equation \ref{eq:11}). The left and right panels are for the cases of different $\eta$.  We also show the radius in terms of the half-SFR radius $R_{\rm 1/2}$ ($= 1.68h_{\rm R}$ for an exponential disk) on the top axis. }
  \label{fig:1}
\end{figure*}

Similarly, substituting Equations \ref{eq:5} and \ref{eq:6} into Equation \ref{eq:3} and setting $\partial{Z}/\partial{t}=0$, we can obtain the steady-state radial differential equation as: 
\begin{equation} \label{eq:7}
    \frac{dZ}{dx} = -\frac{y_{\rm eff}\cdot x}{(1+\eta)e^{x}-(x+1)}, 
\end{equation}
where $y_{\rm eff}$ is the effective yield \citep[also see][]{Wang-21}, defined as
\begin{equation} \label{eq:8}
    y_{\rm eff} = y/(1-R+\lambda). 
\end{equation}
We remind readers that the effective yield $y_{\rm eff}$ defined in this way is different from that in some previous papers \citep[e.g.][]{Edmunds-90, Garnett-02}. 

Combining the boundary condition that the metallicity at the radius of infinity should equal the metallicity of the original inflowing gas ($Z_0$), i.e. $Z(r=\infty)=Z_0$, we can solve the Equation \ref{eq:7} and obtain the steady-state gas-phase metallicity profile as: 
\begin{equation} \label{eq:9}
    Z(r) = - y_{\rm eff}\cdot \ln(1-\frac{x+1}{\eta+1}\cdot e^{-x}) + Z_0
\end{equation}

It should be noted that we can obtain the exact analytic solution of the gas-phase metallicity at the steady-state only under the assumptions of Section \ref{sec:2}. 
In our model, where the exponential profile of the SF disk is treated as an input, both the radial inflow rate and the gas-phase metallicity can then be uniquely determined. 

An important and possibly counter-intuitive consequence of this is that the solution for the metallicity in Equation \ref{eq:9} does {\it not} depend on the radial cold gas distribution, nor (equivalently) on the assumed star formation law, i.e. on any radial dependence of the SFE. 
This is because the radial inflow velocity ($v_{\rm r}$) changes with varying gas surface density when the inflow rate is fixed: 
\begin{equation} \label{eq:10}
    \Phi(r) = 2\pi r \Sigma_{\rm gas} v_{\rm r}. 
\end{equation}
It can be seen from Equation \ref{eq:10} that, for a given annulus within the gas disk, a higher $\Sigma_{\rm gas}$ (equivalent to a lower SFE for a fixed $\Sigma_{\rm SFR}$) corresponds to a lower $v_{\rm r}$.
This slower inflow rate requires more time for the gas to flow across a given annulus, and thus there is more enrichment by the (fixed) star-formation rate. This greater enrichment compensates for the higher $\Sigma_{\rm gas}$, resulting in exactly the same metallicity as the case of lower $\Sigma_{\rm gas}$. 

The inflow velocities in Paper I can reach 50-100 km s$^{-1}$ in the outermost regions of the disk, but this can be reduced if the SFE is lower than assumed (because $\Sigma_{\rm gas}$ is then larger, so the required $v_{\rm r}$ is smaller).  This is broadly consistent with the predictions from hydrodynamical simulations \citep{Trapp-21, SWang-22}, but larger than recent observational analyses indicate \citep{Schmidt-16, DiTeodoro-21}.  In a future paper (E. Wang \& S.J. Lilly 2022, in preparation), we will however investigate the kinematic features of radial gas inflows and discuss the strong degeneracy in kinematic data between radial flows and the effect of warped disks.

Observationally, the gas-phase metallicity is usually shown in logarithmic space, e.g. the Oxygen abundance $\log$(O/H). According to Equation \ref{eq:9}, once we specify any central sink term $\eta$, the shape (i.e. the radial profile in our normalised radius $x$) of the gas-phase metallicity profile in log space, depends {\it only} on the ratio $Z_0/y_{\rm eff}$. 
We show the metallicity profile (solid curves) as a function of normalized radius $x$ (=$r/h_{\rm R}$) for a set of different $Z_0/y_{\rm eff}$ and for two different $\eta$ in Figure \ref{fig:1}. 

As can be seen, at fixed $\eta$, the $\log Z$ profile becomes flatter with increasing $Z_0/y_{\rm eff}$. If the accreting gas is pristine ($Z_0 =0$), the $\log Z$ profile is close to a straight line at $r>h_{\rm R}$, indicating that the $Z(r)$ is close to an exponential function of radius. For $Z_0>0$, the $\log Z$ decreases rapidly at small radii, and become flat at large radii at the value of $\log Z_0$.  Comparing the left and right panels of Figure \ref{fig:1}, the size of the central sink term $\eta$ only changes the shape of $\log Z$ in the innermost regions ($r<h_{\rm R}$). Increasing $\eta$ reduces the cusp of $\log Z$ in the galactic center that is otherwise present for very small $\eta$. 

Equation \ref{eq:9} (or Equation \ref{eq:xx2}) directly reflects the outside-in metal-enrichment of the gas disk as the gas flows inward in our model. At very large radii, the accreted gas has not been significantly enriched by the star formation, and is therefore close to $Z_0$. With flowing inward, the gas is gradually enriched by in-situ star formation on the disk. This process is controlled by two factors, the effective yield of the star formation and the scale-length of $\Sigma_{\rm SFR}$. The $y_{\rm eff}$ determines the strength of metal enrichment, while the $h_{\rm R}$ determines the radial steepness. 

Motivated by this, we provide an approximate simple form for $Z(r)$ with respect to Equation \ref{eq:9},  which is an exponential function plus a constant $Z_0$:
\begin{equation} \label{eq:11}
    Z(r) \simeq a \cdot y_{\rm eff}e^{-bx} + Z_0
\end{equation}
We find that $a =3.0$ and $b=0.86$ gives a very good agreement with the Equation \ref{eq:9} at $r>h_{\rm R}$. We show the curves of Equation \ref{eq:11} in both panels of Figure \ref{fig:1} in the gray dashed lines.   

In fact, the curves from Equation \ref{eq:11} match well those from Equation \ref{eq:9} over the whole range of radius for $\eta=0.05$. Such a big $\eta$ may not be realistic. We note however that the central cusp in $\log Z$ for low $\eta$ could also be reduced by diffusion effects (see Section \ref{sec:3.3}).

\subsection{The diffusion effect} \label{sec:3.3}

\begin{figure}
  \begin{center}
    \epsfig{figure=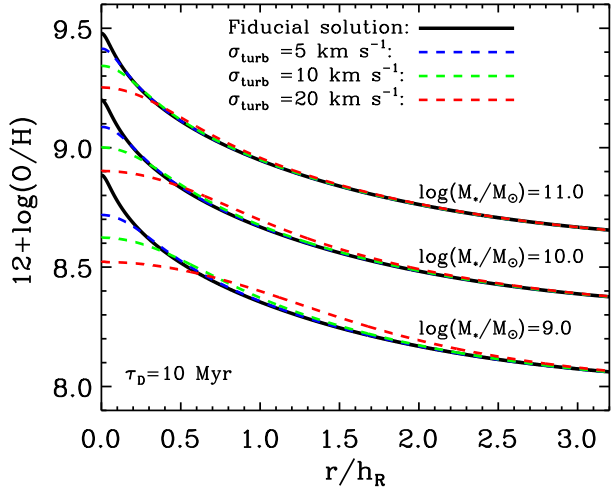,clip=true,width=0.48\textwidth}
   \end{center}
   \caption{The effect of diffusion on the metallicity profile for three typical Main Sequence galaxies of different masses. The colored dashed lines show the numerical solutions of Oxygen abundance for three different $\nu_{\rm D}$, as denoted in the top of the panel. For comparison, we show the metallicity profile without any diffusion as the black solid lines (also see Equation \ref{eq:9}).  }
  \label{fig:8}
\end{figure}

Owing to both stellar winds and supernova explosions, the gas disk can be highly turbulent, especially at high-redshift \citep{Wisnioski-15}. This would lead to radial gas mixing, which would tend to reduce the radial gradients in the Oxygen abundance. In this subsection, we try to quantify this diffusion effect on the gas-phase metallicity profile produced by the modified accretion disk model.  
For a phenomenological description of this process, we can introduce a diffusion term into Equation \ref{eq:2}. Then the continuity equation of metal mass can then be written as: 
\begin{equation} \label{eq:21}
    \Sigma_{\rm gas} \cdot \frac{\partial Z}{\partial t} = \Phi \cdot \frac{\partial Z}{2\pi r \partial r} + y\cdot \Sigma_{\rm SFR} + \frac{\partial}{r\partial r}(\nu_{\rm D}\Sigma_{\rm gas}r\frac{\partial Z}{\partial r}), 
\end{equation}
where $\nu_{\rm D}$ is the diffusion coefficient. The turbulence-driven $\nu_{\rm D}$ is proportional to the squared of the turbulent velocity and the dissipation timescale \citep{Karlsson-13}: 
\begin{equation} \label{eq:22}
\nu_{\rm D} \sim \sigma_{\rm turb}^2 \cdot \tau_{\rm D}. 
\end{equation}
We note that the $\sigma_{\rm turb}$ in Equation \ref{eq:22} reflects the turbulent velocity in the radial direction only, as opposed to the 3-dimensional turbulent velocity, since we are only concerned with the radial dimension in the current work.  

We assume the $\tau_{\rm D}$= 10 Myr \citep{Wada-02, Mac-Low-04}, which roughly corresponds to a turbulent scale length of 100 pc for $\sigma_{\rm turb}=10$ km s$^{-1}$.  Observationally, the $\sigma_{\rm turb}$ can be traced by the velocity dispersion of the CO and HI gas. The typical velocity dispersion in the cold gas is 15-20 km s$^{-1}$ in the inner regions, decreasing to 6-10 km s$^{-1}$ at the outskirts of local SF galaxies \citep[e.g.][]{Boulanger-92, Petric-07, Boomsma-08, Tamburro-09}.  We therefore explore the effect of diffusion for different $\sigma_{\rm turb}$=5, 10 and 20 km s$^{-1}$.  For simplicity, we assume that $\nu_{\rm D}$ is independent of radius.

In order to solve Equation \ref{eq:21}, we consider three typical Main Sequence galaxies of different stellar masses, $M_* = 10^9$, $10^{10}$ and $10^{11}$\msolar\ in the local universe.  Their SFRs, sizes and mass-loading factors follow the assumptions adopted below in Section \ref{sec:4.1} (see Equations \ref{eq:14}, \ref{eq:15}, \ref{eq:16}, and \ref{eq:17}). The Oxygen yield in terms of element number fraction is assumed to be ${\rm O/H}=8.4\times10^{-4}$  with \cite{Chabrier-03} IMF, which corresponds to 0.01 in mass fraction assuming that the Helium contributes one-fourth of the total mass \citep[][]{Vincenzo-16}. The metallicity of the inflowing gas at the outer boundary is assumed to be identical to the $y_{\rm eff}$ (see Section \ref{sec:5}). For each stellar mass, we obtain the numerical solutions of $Z$ at equilibrium with three different $\nu_{\rm D}$, shown in Figure \ref{fig:8}. 

Figure \ref{fig:8} basically gives an impression of the potential significance of the diffusion effect. As shown, turbulence-driven diffusion can indeed flatten the gas-phase metallicity profile with respect to the fiducial solution (Equation \ref{eq:9}) in the inner regions of galaxies ($<$0.5-1.0$h_{\rm R}$).  As expected, this flattening effect becomes more significant with increasing $\nu_{\rm D}$. At given $\nu_{\rm D}$, the flattening effect is more significant for less massive galaxies, because of their smaller sizes.  
We conclude that the diffusion effect can reduce the gradients of gas-phase metallicity within 0.5-1$h_{\rm R}$ in the our model, and can completely eliminate the cusp-structure in $\log Z$ that can otherwise appear at the galactic center.

\subsection{The global gas-phase metallicity} \label{sec:3.4}
It is also of interest to calculate an average metallicity of the gas in the galaxy.  This is easily done analytically when $\eta \sim 0.0$.  We calculate the SFR-weighted average metallicity. This is for two reasons. First, we do not want to have to assume any particular gas surface density (i.e. SFE) profile here -- recall that it is the star-formation profile which is input to the model. Second, the observed Oxygen abundance is usually measured using line ratios of emission lines from HII regions, which will approximate an SFR-weighted measurement of the gas metallicity. 

The SFR-weighted metallicity can be written as 
\begin{equation} \label{eq:13}
\begin{split}
Z_{\rm tot} 
& = \frac{\int_0^\infty Z(r)\cdot 2\pi r\Sigma_{\rm SFR}(r)dr}{\int_0^\infty 2\pi r\Sigma_{\rm SFR}(r)dr}  \\
& = y_{\rm eff}\cdot \int_0^\infty (x-\ln(e^x-x-1))\cdot xe^{-x}dx + Z_0  \\
& = y_{\rm eff} +Z_0
\end{split}
\end{equation}

This SFR-weighted average metallicity $Z_{\rm tot}$ has a remarkably simple form. It is just the metallicity of the inflowing gas at the outer boundary of the disk, plus the effective yield within the galaxy at the epoch in question. This is the same as the prediction in the simple
single-chamber gas-regulator system \citep[e.g.][]{Sanchez-Almeida-14, Lilly-13, Wang-21} fed at constant rate.  This is essentially because the SFR-weighted metallicity is the same as the average gas mass-weighted metallicity if we adopt a constant SFE, as usually assumed in the simple gas-regulator system.  However, we stress that the Equation \ref{eq:13} for the SFR-weighted metallicity holds in our accretion disk system {\it independent} of the constancy or otherwise of the SFE.  The global SFR-weighted metallicity of Equation \ref{eq:13} is a general and robust prediction, which is free from the effects of diffusion and other assumptions about the radial dependence of the relevant parameters.   

We note that Equation \ref{eq:13} gives the predicted SFR-weighted metallicity for the whole gas disk, while the observed ``integrated" Oxygen abundance may in practice only have been measured from some more limited area of the galaxies. For instance, 
the gas-phase metallicities of large numbers of galaxies in the SDSS (The Sloan Digital Sky Survey) have been measured within the central 3 arcsec fibre area \citep{Tremonti-04}.  Therefore, to present a fair comparison with observations, one should calculate the model-predicted metallicity with Equation \ref{eq:13} but integrating only out to the corresponding radius of the observations.

\section{The model predictions in the context of evolution}  \label{sec:4} 

\begin{figure*}
  \begin{center}
    \epsfig{figure=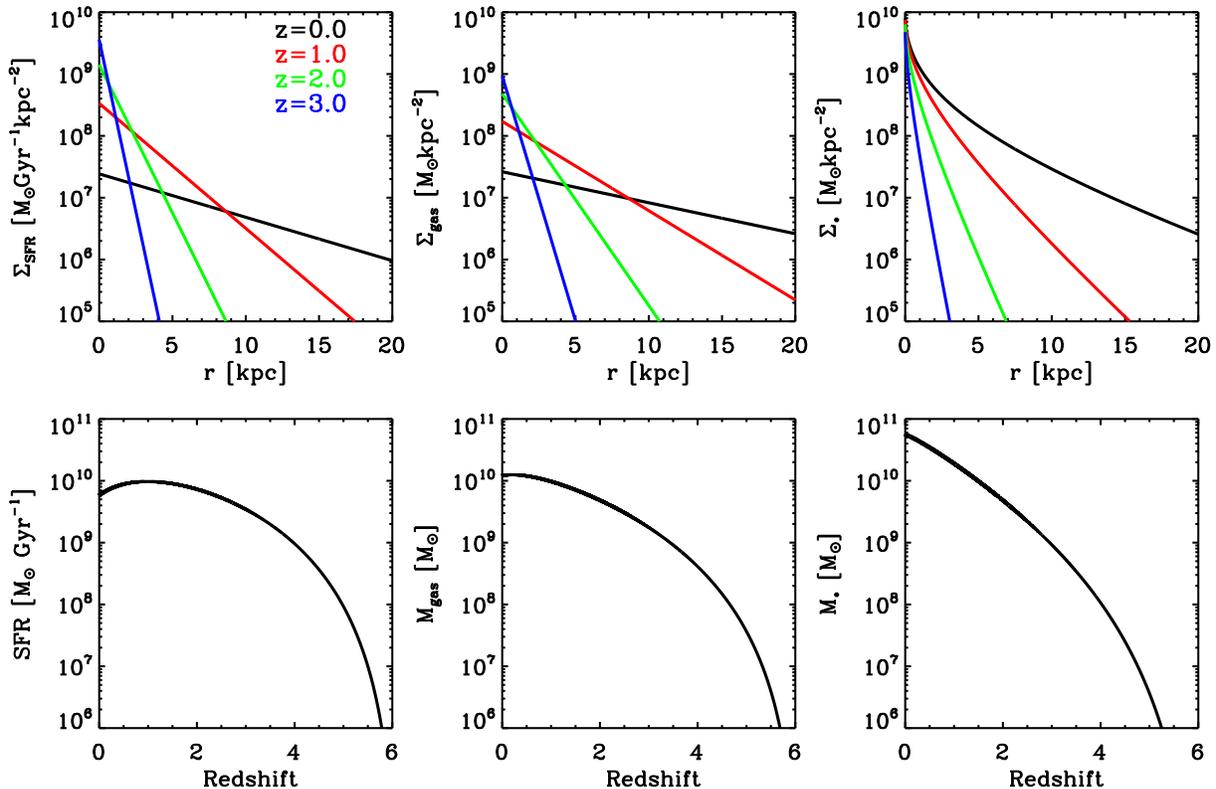,clip=true,width=0.9\textwidth}
    \end{center}
  \caption{Illustration of the input condition of $\Sigma_{\rm SFR}$ and star formation law for the evolution model of galactic disks. The top panels shows the evolution of $\Sigma_{\rm SFR}$, $\Sigma_{\rm gas}$ and $\Sigma_{*}$ for a typical MS galaxy ($M_*(z=0) = 10^{10.75}$\msolar), while the bottom panels show the evolution of the integrated SFR, cold gas mass and stellar mass for this same galaxy.  } 
  \label{fig:2}
\end{figure*}

In Section \ref{sec:3}, we obtained the radial profile of the co-planar inflow rate and the gas-phase metallicity, under the assumption of a steady-state equilibrium (see Equation \ref{eq:4}) in which a steady inflow of gas is progressively consumed by star-formation (or lost as a wind).  However, it is unclear whether the gas disk of galaxies is exactly in such a (quasi-)equilibrium state or not. Strong evolution in size and SFR has been found for SF galaxies over cosmological timescales.  Based on many spatially resolved spectroscopic surveys \citep[e.g.][]{Croom-12, Bundy-15}, SF galaxies are found to grow inside-out, as indicated by the radial distribution of their stellar populations \citep[e.g.][]{Perez-13, Li-15, Goddard-17, Rowlands-18, Wang-18}. This is also consistent with the fact that the size-mass relation of SF galaxies changes with time.  Star-forming galaxies were smaller, at a fixed mass, at earlier epochs \citep[e.g.][]{Toft-07, Williams-10, Newman-12, vanderWel-14}.  Therefore, the assumption of a steady-state in Equation \ref{eq:4} may be not fully applicable. 

In this subsection, we set up a heuristic time-varying model based on Equation \ref{eq:1} and \ref{eq:2}.  The evolving model we construct is intentionally very simple. However, it is enough to answer the question as to what the effect of size and SFR evolution is in determining the co-planar inflow rate and gas-phase metallicity, with respect to the steady-state solution discussed above.  

\subsection{Settings of the time-evolving model} \label{sec:4.1}


The time-evolving model works as follows. First, we start from a SF disk galaxy at redshift of 6 with a small seed stellar mass and an exponential gas disk. 
Then we let it evolve in stellar mass following the evolution of the star formation main sequence (SFMS). The evolution of SFMS is obtained from observations \citep[e.g.][]{Stark-13, Speagle-14}. In each time step, we assume that the newly formed stars are exponentially distributed within the disk, with a scale-length that evolves with time. The evolution of the disk scale-length can be constrained by matching the observed time-dependent mass-size relation of SF galaxies. Thus far, the model is conceptually similar to that in \cite{Lilly-16}.  

Based on this evolving $\Sigma_{\rm SFR}(r,t)$, the instantaneous gas surface density $\Sigma_{\rm gas}(r,t)$ is obtained by assuming a star formation law.  Then, based on Equation \ref{eq:1} and Equation \ref{eq:2}, we can calculate the required co-planar inflow rate and gas-phase metallicity at each time-step in this evolving model, with the usual assumptions of a mass-loading factor and the metallicity of inflowing gas.    

Specifically, we assume that the evolution of the sSFR, defined as SFR/$M_*$, follows the formula: 
\begin{equation} \label{eq:14}
  {\rm sSFR}(M_*,z) = \frac{0.07}{1-R}\times  (\frac{M_*}{3\times 10^{10}M_{\odot}})^{-0.2}\times (1+z)^2 \ {\rm Gyr^{-1}}
\end{equation}
Throughout this work, the stellar mass is defined as the mass of living stars.  This slightly sub-linear SFMS and its evolution with redshift is broadly consistent with many observational results \citep[e.g.][]{Pannella-09, Stark-13, Speagle-14}. 
The evolution of the scale-length of $\Sigma_{\rm SFR}$ is assumed to be: 
\begin{equation} \label{eq:15}
    h_{\rm R}(M_*,z) = 5 \times (\frac{M_*}{3\times 10^{10}M_{\odot}})^{1/3}\times (1+z)^{-1} \ {\rm kpc}
\end{equation}
Both Equation \ref{eq:14} and \ref{eq:15} are taken from \cite{Lilly-16}, who have shown that these two equations work well in reproducing  the mass-size relation of galaxies.  


We assume a time-independent star formation law, taken from \cite{Kennicutt-98}: 
\begin{equation} \label{eq:16}
    \Sigma_{\rm SFR} = 2.5\times 10^{-4} \cdot (\frac{\Sigma_{\rm gas}}{1\ {\rm M}_{\odot}{\rm pc}^{-2}})^{1.4} \ \ {\rm M_{\odot}yr^{-1}kpc^{-2}}
\end{equation}
The mass-loading factor cannot usually be directly measured from observations. Based on the Feedback in Realistic Environments (FIRE) simulations, \cite{Muratov-15} found that the $M_*$-$\lambda$ relation can be well described by an power law relation and is nearly redshift-independent:  
\begin{equation} \label{eq:17}
 \lambda= 3.6\times (\frac{M_*}{10^{10}{\rm M}_{\odot}})^{-0.35}
\end{equation}
We adopt this time-independent $M_* - \lambda$ relation in the time-evolving model. The Equation \ref{eq:17} may overestimate the mass-loading at high stellar mass end for galaxies in the local universe  \cite[see][]{Hayward-17}. 
We note that the mass-loading factor taken from \cite{Muratov-15} is not exactly the same as we defined in Section \ref{sec:2}, where we combine the outflow with any ex-planar inflow to define an effective outflow. However, the effective $\lambda$ is close to the real mass-loading factor when the ex-planar inflow is negligible, as suggested from the simulations \citep{Peroux-20, Trapp-21, Hafen-22}.
The metallicity of inflowing gas likely becomes higher during the evolution, because the CGM becomes progressively metal-enriched with time due to the wind-outflow from the host galaxy.  We assume the $Z_0$ is equal to $y_{\rm eff}$ at each time-step, motivated by the result in Section \ref{sec:5}. In the time-evolving model, we calculate the gas-phase metallicity in terms of the Oxygen element fraction, rather than the mass fraction. The Oxygen yield in terms of element number fraction is assumed to be $8.4\times10^{-4}$ \citep[][]{Henry-00}, the same as the one assumed in Section \ref{sec:3.3}. The $\eta$ is assumed to be 0.001 and to not change with time. 

Needless to say, not all of these assumptions are necessarily fully realistic. We emphasize that our purpose in this subsection is to examine the effect of size and SFR evolution on the steady-state solution obtained earlier, rather than to present a fully realistic evolution of gas-phase metallicity. 

For this purpose, we construct the evolution model for a typical galaxy, which has stayed on the SFMS all of its life and which has a Milky-Way mass at redshift zero ($M_*=10^{10.75}{\rm M}_{\odot}$ at $z=0$). Figure \ref{fig:2} shows the evolution of $\Sigma_{\rm SFR}$, $\Sigma_{\rm gas}$ and $\Sigma_*$ for this object at four different redshifts, as well as the evolution of the integrated quantities. 
In the next subsection, we will solve the Equation \ref{eq:1} and \ref{eq:2} at each time-step numerically with the above settings, and compare the solutions with the steady-state solutions.   

\subsection{The radial inflow rate and metallicity in the time-evolving model}   \label{sec:4.2}

\begin{figure*}
  \begin{center}
    \epsfig{figure=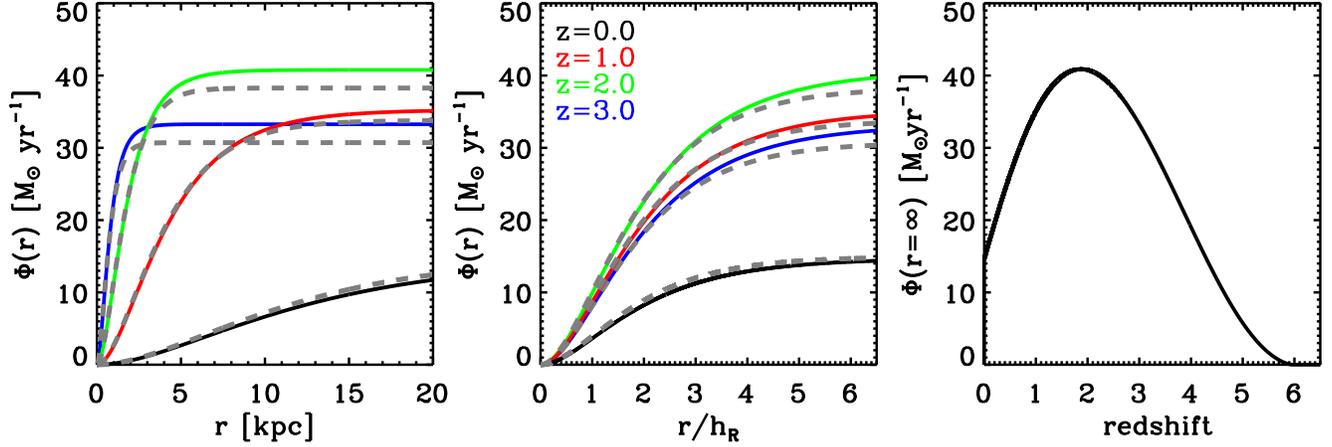,clip=true,width=0.99\textwidth}
    \end{center}
  \caption{The numerical solution of the radial inflow rate at different epochs. For comparison, we show the 
  the solution of the radial inflow rate in the steady-state at the same epochs as dashed gray curves. The left and middle panels show the inflow rate as a function of physical radius and the radius scaled with the scale-length, while the right panel show the evolution of total inflow rate. }
  \label{fig:3}
\end{figure*}

\begin{figure*}
  \begin{center}
    \epsfig{figure=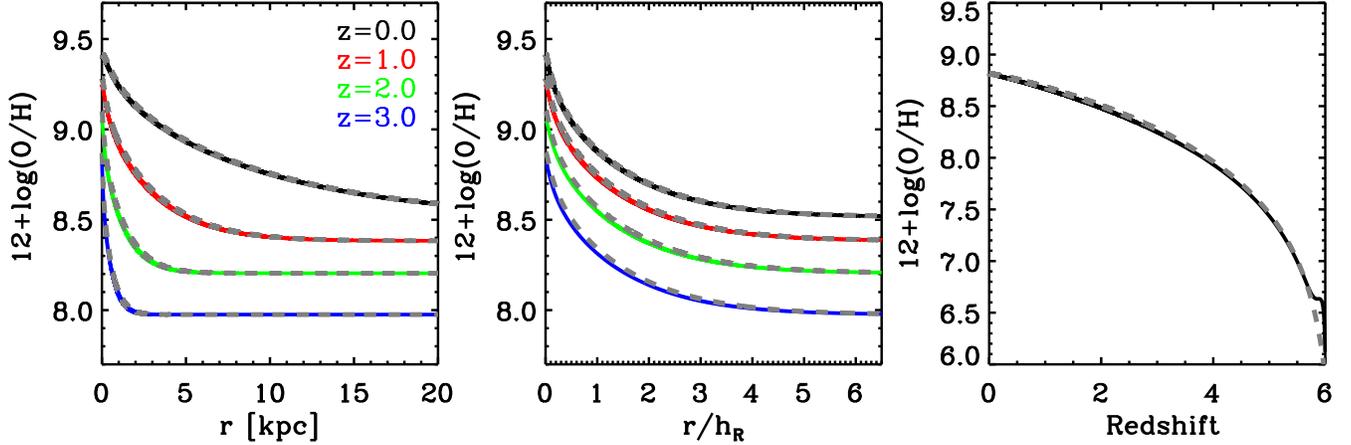,clip=true,width=0.99\textwidth}
    \end{center}
  \caption{The numerical solutions for the evolving gas-phase metallicity at different epochs. For comparison, we show the 
  the steady-state solutions at the same epochs as dashed gray curves. These are excellent representations of the state of the evolving model. The left and middle panels show the metallicity as a function of the physical radius and the radius scaled with the scale-length, while the right panel show the evolution of global metallicity, computed as the SFR-weighted gas-phase metallicity. }
  \label{fig:4}
\end{figure*}

As illustrated in Figure \ref{fig:2}, the evolution of $\Sigma_{\rm SFR}$ and $\Sigma_{\rm gas}$ for the typical Milky-Way mass galaxy are fixed. Therefore, in the framework of our model, we can calculate the required inflow rate numerically as follows. The length of time-step is denoted as $\Delta t$, and the length of radial step is denoted as $\Delta r$. 
Based on Equation \ref{eq:1}, the inflow rate at $i$th time step and $j$th radial step, denoted as $\Phi(r_j, t_i)$, can be numerically computed as: 
\begin{equation} \label{eq:18}
\begin{split}
  \Phi(r_j, t_i) = & \Phi(r_{j-1}, t_i) + 2\pi r_j \Delta r \cdot (1-R+\lambda(t_i))\Sigma_{\rm SFR}(r_j,t_i)  \\
  & + 2\pi r_j \Delta r \cdot  (\Sigma_{\rm gas}(r_{j}, t_i) - \Sigma_{\rm gas}(r_{j}, t_{i-1})) / \Delta t, 
\end{split}
\end{equation}
where $\lambda(t_i)$ is determined by $M_*(t_i)$ following the Equation \ref{eq:17}.  Then we can obtain the $\Phi(r_j,t_i)$ with the boundary condition of $\Phi(r=0,t_i)$. 
Differently from the steady-state solution, we here include the change of $\Sigma_{\rm gas}$ (the third term in the right-hand of Equation \ref{eq:18}). 
After obtaining $\Phi(r_j,t_i)$, we can further obtain the $Z(r_j, t_i)$ based on Equation \ref{eq:3} using the backward Euler method: 
\begin{equation} \label{eq:19}
\begin{split}
    & \Sigma_{\rm gas}(r_j, t_i)\cdot \frac{Z(r_j,t_i)-Z(r_j, t_{i-1})}{\Delta t} = \\
    & \Phi(r_j, t_i) \frac{Z(r_{j+1}, t_i) - Z(r_j, t_i)}{2\pi r_j \Delta r} + y\cdot \Sigma_{\rm SFR}(r_j, t_i)
\end{split}
\end{equation}
We can then compute $Z(r_j, t_i)$ from Equation \ref{eq:19} with the boundary condition $Z(r=\infty, t_i) = Z_0(t_i)$, and the initial condition $Z(r,t=0.0)$ = $Z_0(t=0.0)$. The backward Euler method enables us to obtain a convergent solution for the gas-phase metallicity. Equation \ref{eq:19} takes into account the cosmic evolution of metallicity, with respect to the steady-state. 

The left and middle panels of Figure \ref{fig:3} show the numerical solutions (solid curves) of $\Phi(r)$ at different epochs for the typical galaxy (see Figure \ref{fig:2}), with the radius in kpc and $h_{\rm R}$.  For comparison, each of the solid curves is associated with a gray dashed curve, which is the steady-state solution of $\Phi(r)$ at the corresponding epoch obtained previously using Equation \ref{eq:6}.  The evolution of total inflow rate, $\Phi(r=\infty)$, is shown in the right-most panel of Figure \ref{fig:3}. 

As shown, the total inflow rate first gradually increases, and then decreases after a peak at z$\sim$1.7. Since $\Phi(r)$ is a monotonically increasing function with radius, the $\Phi(r=\infty)$ determines the overall strength of $\Phi(r)$, and the scale-length of $\Sigma_{\rm SFR}$ mainly determines the radial gradient of $\Phi(r)$.  

Comparing with the steady-state solutions (Equation \ref{eq:6}), scaling the radius by $h_{\rm R}$, we find that generally the steady-state solutions obtained earlier agree rather well with the time-evolving solutions out to at least 4$h_{\rm R}$
at different epochs. The main difference is that the $\Phi(r)$ for the evolving solution is slightly higher at high redshift than the steady-state predication at larger radii. 
This is due to the fact that the galactic disk is growing its size.  At $z=0$, the time-evolving solution appears to slightly lower than the steady-state calculation, because the total SFR is gradually decreasing at the current epoch (see Figure \ref{fig:2}).

The left and middle panels of Figure \ref{fig:4} show the numerical solutions of $Z$ at different epochs as the solid curves with radius in kpc or in $h_{\rm R}$, together with the steady-state solution of $Z$ at the corresponding epoch (see Equation \ref{eq:9}).  The right panel of Figure \ref{fig:4} shows the cosmic evolution of the SFR-weighted metallicity for the modelled galaxy. 

It is striking that the time-evolving solution of $Z$ is nearly identical to the steady-state solution. This indicates that we can apply the steady-state solutions for the radial dependence of the gas-phase metallicity derived in Equation \ref{eq:9} also in an evolving framework, provided that the galaxies are evolving more or less as expected. 
Although the SFR and size of galaxies evidently do evolve significantly with time, we conclude that their gas disks can be treated as quasi-equilibrium steady-state systems at any single epoch, at least in terms of the co-planar inflow rate and the gas-phase metallicity.   

The radial gradient of the metallicity strongly depends on the choice of x-axis, i.e. dex/kpc or dex/$h_{\rm R}$. According to Equation \ref{eq:9} or Equation \ref{eq:11},  for a given $y_{\rm eff}/Z_0$, the gradient of the gas-phase metallicity in dex/kpc {\it only} depends on the scale-length of galactic disk. Our simple model predicts that smaller galaxies tend to have a steeper radial gradient of $Z$, indicating that high-redshift galaxies have steeper metallicity gradients when measured in dex/kpc (see Figure \ref{fig:4}).  This appears to be consistent with the findings in \cite{Jones-13}, where they find some high-redshift galaxies show extremely steep metallicity gradients measured in dex/kpc \citep[also see][]{Wang-17}.  However, we note that some of high-redshift SF galaxies may not have well-defined disk structure, due to heavy inflow and frequent mergers. 
Based on the FIRE simulation, \cite{Ma-17} found that strong negative metallicity gradients are only found in galaxies with a rotating disk, and strongly perturbed galaxies with little rotation (for which the accretion disk model is unlikely to be valid) always show small gradients. 
Anyway, we argue that it is more revealing to measure the gradient of gas-phase metallicity in dex/$h_{\rm R}$ than in dex/kpc, if the co-planar inflow is significant as assumed here. 

\section{Application to the observed radial profiles of gas-phase metallicity} \label{sec:5}

In Section \ref{sec:3}, we have provided the metallicity profiles of galactic gas disks in a simple analytic form under the modified accretion disk model assumed to be in a steady-state. In Section \ref{sec:4} we showed that these ``steady-state" solutions were in fact also applicable to galaxies that were evolving in both size and SFR at the rates indicated by observations.   In this section, we apply this steady-state analytic formula to the observed profiles of gas-phase metallicity in galaxies, and examine whether it is able to match the observational results. 


\begin{figure*}
  \begin{center}
    \epsfig{figure=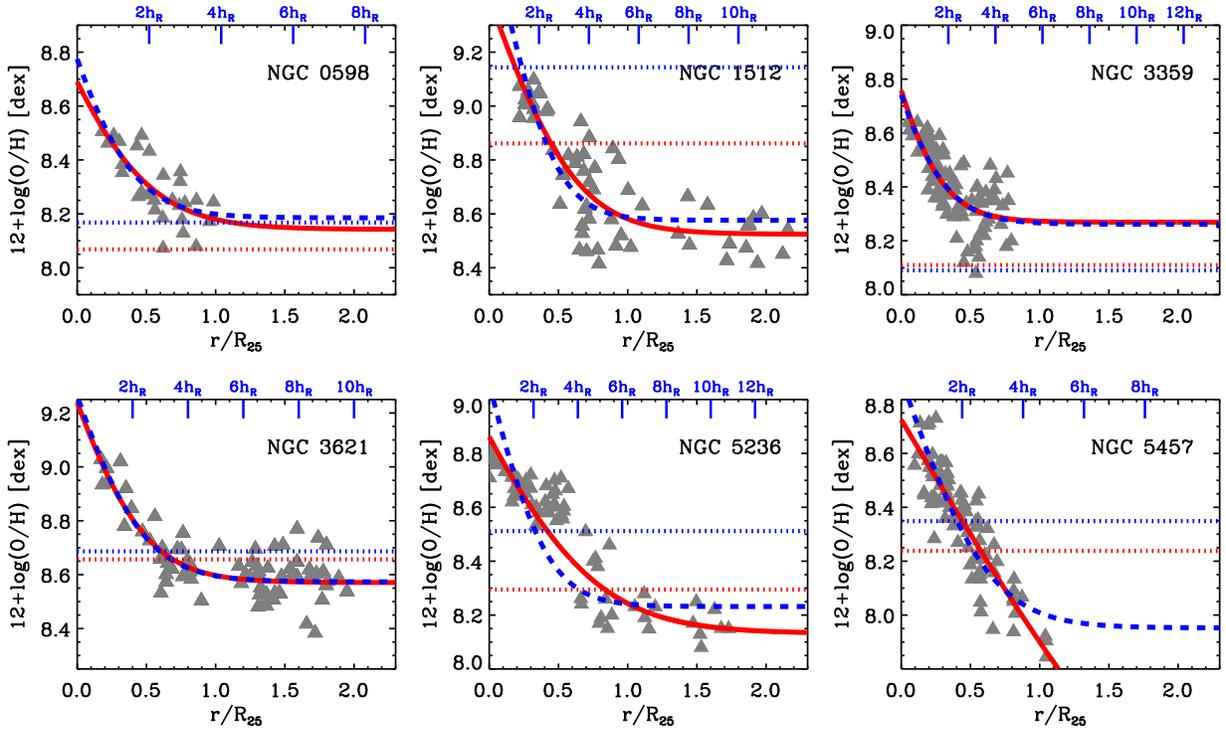,clip=true,width=0.92\textwidth}
    \end{center}
  \caption{The gas-phase metallicity profile for the six individual galaxies examined. The data points are extracted from the literature (see the last column of Table \ref{tab:1}). 
  The observed data points are fitted by Equation \ref{eq:11}, both with $h_{\rm R}$ as a free parameter (red curves) and as a fixed parameter obtained from independent observations (blue dashed curves). In each panel, the two horizontal dotted lines indicate the fitted values of $y_{\rm eff}$ for the two fitting approaches.   
  We also show the radius in terms of $h_{\rm R}$ (from the observation) at the top of each panel.  }
  \label{fig:5}
\end{figure*}

\begin{table*}[ht]
\renewcommand\arraystretch{1.5}
\begin{center}
\caption{The fitted parameters of metallicity profile for the six galaxies in Figure \ref{fig:5} \label{tab:1}}
\begin{tabular}{@{}lrrrrrrrrr@{}}
\tableline
\tableline
                { }   &        { }   &  { }     & Fit\_para  & (Free $h_{\rm R}$)  &   { }   & Fit\_para  &  (Fixed $h_{\rm R}$) &  { } \\      
\cmidrule(lr){4-5}
\cmidrule(lr){7-8}

 Name               &   O/H-Indicator    &   { }  &  $Z_0/y_{\rm eff}$  & $h_{\rm R}/R_{\rm 25}$ & { }  &  $Z_0/y_{\rm eff}$  & $h_{\rm R}/R_{\rm 25}$  &  References  \\
\tableline
   NGC 0598                     &   T$_{\rm e}$-method        & { }    &  1.18     &   0.26      & { }    &  1.04    &  0.19$^a$    & {\scriptsize \cite{Crockett-06, Magrini-07}}   \\
 NGC 1512                     &   R$_{\rm 23}$         & { }    &  0.46      &   0.22      & { }    & 0.27   &  0.14$^a$  & {\scriptsize \cite{Bresolin-12}} \\
 NGC 3359                   &       O3N2     & { }    &  1.44     &   0.16      & { }    & 1.48    &   0.17$^b$   & {\scriptsize \cite{Martin-95, Scarano-13}}  \\
 NGC 3621                 &     R$_{\rm 23}$  & { }    &  0.82     &   0.21      & { }    & 0.77    &   0.20$^c$   & {\scriptsize \cite{Bresolin-12}}  \\
 NGC 5236                 &   N2O2      & { }    &  0.69     &   0.32      & { }    & 0.52    &    0.16$^c$   & {\scriptsize \cite{Bresolin-09, Scarano-13}} \\
 NGC 5457                 &   T$_{\rm e}$-method & { }    &  0.06    &   0.42      & { }    & 0.40    &    0.22\tablenotemark{c}   & {\scriptsize \cite{Kennicutt-03, Croxall-16}} \\
\tableline
\tableline
\end{tabular}
\\

{\raggedright  $^a$ The scale-length of 70$\mu$m-band image \citep{Rieke-04}. \\ }
{\raggedright $^b$ The scale-length of H$\alpha$ emission \citep{Rozas-00}. \\}
{\raggedright $^c$ The scale-length of $\Sigma_{\rm SFR}$ obtained from the combined UV+IR images \citep{Casasola-17}. \\}
\end{center}
\end{table*}


We apply Equation \ref{eq:11} to six nearby galaxies: NGC 0598, NGC 1512, NGC 3359, NGC 3621, NGC 5236, and NGC 5457. The gas-phase metallicity of HII regions for these six galaxies are extracted from the literature (see Table \ref{tab:1}). These six galaxies are selected because they have well measured metallicity profiles up to the radius of $R_{\rm 25}$ or more, corresponding to $\sim$4 scale-lengths of their stellar disks \citep{Hakobyan-09}. 
The large range in radius ensures the inclusion of radii where we should see the flattening, and therefore allows us in principle to obtain $Z_0$, $y_{\rm eff}$ and $h_{\rm R}$ by fitting the metallicity profile with Equation \ref{eq:11}. We can then compare the fitted $h_{\rm R}$ with the observed $h_{\rm R}$ which is directly measured from the $\Sigma_{\rm SFR}$ profile.  This latter consistency check provides validation of our simple model.   


Figure \ref{fig:5} shows the radial profiles of 12+log(O/H) for the six galaxies with the radius scaled by $R_{\rm 25}$, which is defined as the 25 mag arcsec$^{-2}$ B-band isophote. Despite the large range of methods for obtaining the Oxygen abundance from different observations by different authors, a pronounced flattening is seen in the metallicity profile at radii of 0.6-1.0$R_{\rm 25}$ for five of the six galaxies. The only exception is NGC 5457. This evidence for the predicted flattening appears to be in good agreement with our model prediction (see Figure \ref{fig:3}). 

The flattening of Oxygen abundance in the outer regions of disk has been known for decades \citep[e.g.][]{Mishurov-02, Scarano-13}.  \cite{Scarano-13} found that the flattening radius is comparable to the corotation radius, and interpreted the flattening of Oxygen abundance as a consequence of long-lived spiral structures, where the SFR depends on the distance to the corotation radius. However, the radial profiles of $\Sigma_{\rm SFR}$ (or $\Sigma_{\rm gas}$) do not appear to break for nearby disk galaxies at the corotation radii \citep{Casasola-17}, which appears to be inconsistent with the above explanation. In this paper, we argue that the flattening of gas-phase metallicity at the outer disk is a natural consequence of the co-planar inflow, and is produced by the metallicity floor of the inflowing gas. 
The typical metallicity of the floor is at relatively high levels, 0.3-0.5 $Z_{\odot}$ \citep[e.g.][]{Bresolin-12, Sanchez-14}.  Previous work has also ascribed the flattening of the $Z_{\rm gas}(r)$ profile in the outer disk to the accretion of pre-enriched materials \citep[][and references therein]{Maiolino-19}.

We then perform fits to the observed data points in Figure \ref{fig:5} using Equation \ref{eq:11}, with either $h_{\rm R}$ a free parameter or as a fixed quantity obtained from other observations such as the H$\alpha$ profile (see Table \ref{tab:1}).  
We obtain the $h_{\rm R}$ of these galaxies from the literature \citep{Rozas-00, Casasola-17}, based on either the H$\alpha$ emission or the combined UV+IR images (see details in Table \ref{tab:1}), except for NGC 0598 and NGC 1512. For these two galaxies, we did not find the scalelength of $\Sigma_{\rm SFR}$ from the literature. We therefore measure the scalelength for these two galaxies from the 70$\mu$m-band image \citep{Rieke-04}, which is believed to be a good tracer of star formation \citep{Calzetti-10}.  
In each panel of Figure \ref{fig:5}, the red line is the best-fit profile with free $h_{\rm R}$, and the blue line is the best-fit with the imposed $h_{\rm R}$.  The fitting parameters are listed in Table \ref{tab:1}.

Both fits generally match the data points well, regardless of whether $h_{\rm R}$ was fixed or not. Furthermore, for four of the galaxies (NGC 0598, NGC 1512, NGC 3359 and NGC 3621) the best-fit blue and red curves match well with each other, indicating that the fitted values of $h_{\rm R}$ are comparable to the $h_{\rm R}$ obtained from independent observations. This indicates that our simple model works very well to explain the profile of gas-phase metallicity, at least for these four galaxies. There was no guarantee at all that the fitted $h_{\rm R}$  would match the independently observed $h_{\rm R}$. 

There is a range of fitted $Z_0/y_{\rm eff}$ for these four galaxies, with values $0.3 < Z_0/y_{\rm eff} < 1.5$. The mean is close to $Z_0/y_{\rm eff} \sim 1$ irrespective of whether $h_{\rm R}$ is constrained or not, and this value was therefore adopted in the time-evolving model discussed in Section \ref{sec:4}. 

For two galaxies, NGC 5236 and NGC 5457, the fitted $h_{\rm R}$ is nearly twice the observed value, leading to significant differences between the blue and red curves in the relevant panels of Figure \ref{fig:5}.  The Oxygen abundance profile of NGC 5236 cannot be simply characterized by Equation \ref{eq:11}, because the gas-phase metallicity abruptly decreases at 0.7R$_{\rm 25}$ \citep{Bresolin-09, Scarano-13}. However, as shown in figure 4 of  \cite{Bresolin-09}, this abrupt decrease is less pronounced or even disappears when other metallicity indicators are used. The Oxygen abundance profile of NGC 5457 shown here, does not show any clear flattening out to 1.2$R_{\rm 25}$. However, \cite{Hu-18} have found that a flattening is shown at $\sim$0.8$R_{\rm 25}$ 
(consistent with blue curve) when adopting the metallicity indicator from \cite{Kobulnicky-04}. This serves as a reminder that different methods of measuring Oxygen abundance may play a significant role in determining the detailed features of the profile. 

Our simple model as described by our analytic formula can evidently describe the observed profiles of gas-phase metallicity very well, at least for a majority of galaxies, if not all, provided that the inflowing material (at least at the current epoch) is significantly enriched.  At the very least, we can conclude that the detailed metallicity data on these galaxies certainly does not rule out the modified accretion disk model for galactic disks.   

In principle, we could apply our simple model to the large body of integral-field spectroscopy (IFS) data on nearby galaxies, such as MaNGA \citep[][]{Bundy-15}, CALIFA \citep[][]{Sanchez-12} and SAMI \citep[][]{Croom-12}. However, there are several caveats to comparing the IFS data to the model.  The main concern is that the radial coverage of galaxies in these surveys is quite limited for our model.  As discussed early in this section, the observed metallicity profile for a large range of radius ($\sim R_{25}$ in Figure \ref{fig:5}) is needed to obtain the reasonable fitting parameters, due to the degeneracy of $h_{\rm R}$ and $Z_0/y_{\rm eff}$. 

The measured gradients of Oxygen abundance for the galaxies of these IFS surveys appear to be smaller than those shown in Figure \ref{fig:5} for the six nearby galaxies \citep[e.g.][]{Sanchez-14, Belfiore-19}.  This inconsistency of the observational results can be due to a number of reasons. First, the six nearby galaxies in Table \ref{tab:1} may not be a good representative of the whole SF population due to the limitation of sample size.  Second, these IFS surveys have surely poor spatial resolution than the six nearby galaxies.  \cite{Mast-14} investigated the effect of resolution for the measurement of the metallicity gradients of nearby galaxies, and found that increasing the pixel size could significantly reduce the measured metallicity gradient. In addition, the different coverages and different adoptions of metallicity indicators may also play a role. These issues emphasise the need to map the oxygen abundance to much larger radii ($R_{\rm 25}$ or more) in much larger samples of nearby galaxies. 

According to our steady-state model (see Equation \ref{eq:9}), the flatter profiles of gas-phase metallicity would result in higher values of $Z_0/y_{\rm eff}$.  For instance, the median gradient of log(O/H) for MaNGA SF galaxies is $\sim-0.1$ dex/$h_{\rm R}$ with adopting the {\tt N2H2H$\alpha$} metallicity indicator \citep{Dopita-16, Wang-21},  corresponding to $Z_0\sim 2y_{\rm eff}$ which is quite high. However, our idealized model does not include a number of processes, such as diffusion  (see Figure \ref{fig:8}), the mixing of ex-planar gas at different radii, and any outflow induced by the central massive black hole (see Figure \ref{fig:1} for the effect of $\eta$). These processes could all potentially flatten the metallicity gradient.

\section{Discussion} \label{sec:6}

\subsection{The physical implication of $Z_0/y_{\rm eff}$} \label{sec:6.1}

In this work, we show that the radial gradients of the gas-phase metallicity profile are determined by the scalelength of $\Sigma_{\rm SFR}$ and $Z_0/y_{\rm eff}$, under the modified accretion disk framework.  If the gradients are measured in dex/$h_{\rm R}$, $Z_0/y_{\rm eff}$ is then the only factor to determine the overall shape of $\log Z(r)$.  After applying the model to the six nearby galaxies, we obtain the mean $Z_0/y_{\rm eff}$ of $\sim$1.  In addition to this, we will in this subsection discuss the physical meaning of the value of $Z_0/y_{\rm eff}$.   

According to Equation \ref{eq:13} in Section \ref{sec:3.4}, if the inflowing gas at the outer boundary of the disk is pristine (i.e. $Z_0\sim$0), then the average (SFR-weighted) metallicity of the gas in the disk is just $y_{\rm eff}$, and this is also the average metallicity of the outflowing gas since the outflowing mass is also (in our assumption) scaling with the local SFR. We could imagine that the outflowing gas is later accreted as inflowing gas, again at the outer boundary of the disk, in a cyclical loop.  The metallicity of the inflowing gas $Z_0$ would then be the same as, or less than, $\sim y_{\rm eff}$, depending on the amount of any dilution with pristine gas in the halo.  If it is assumed that the metals in the inflowing gas originated only from the previous outflows of the host galaxies, then the ratio $Z_0/y_{\rm eff}$ formally gives a lower limit to the number of times that gas has cycled through the system. 

If the above statement is true, then the SFE of the system may play an indirect role in determining the metallicity of the inflow gas $Z_0$, even though the SFE does not appear in our steady-state solutions of inflow rate and metallicity (see Equations \ref{eq:xx1}, \ref{eq:xx11} and \ref{eq:xx2}). Galaxies with a higher SFE might be expected to have more recycling, i.e. more cycles in and out of the galaxy in their lifetime, because of the shorter time that gas spends in the galaxy as a result of the shorter gas-depletion timescale. This would result in a larger $Z_0/y_{\rm eff}$. Therefore, we might expect that galaxies with higher SFE could have flatter metallicity profiles and higher global metallicities, than their counterparts of the same mass with lower SFE.

\subsection{The strengths and limitations of our model} \label{sec:6.2}

The basis of the present work is that the gas accretion is dominated by the co-planar radial inflow, and the wind-loading outflow is bipolar and perpendicular to the disk, as found in the hydrodynamical simulations \citep[e.g.][]{Peroux-20, Trapp-21}.  This gives strong constrains on the evolution of the gas disk and the chemical enrichment histories.   
Together with the fact that $\Sigma_{\rm SFR}$ is typically in exponential form for SF galaxies \citep{Wyder-09, Casasola-17, Wang-19}, we are then able to give analytic solutions for the expected gas-phase metallicity profile (see Equation \ref{eq:9}).  This analytic solution is very simple and contains very few free parameters, which furthermore have clear physical meaning. In addition, the predicted metallicity profile is free from the assumed SFE.  Based on our simple model, the negative radial gradient of the gas-phase metallicity is primarily due to the fact that the inflowing gas has been continuously enriched by star formation activities when moving toward the disk center.  

We then apply this analytic solution to the observed profiles of Oxygen abundance for the six nearby galaxies.  By fitting the observed metallicity profile, we can obtain three parameters: the scalelength of $\Sigma_{\rm SFR}$ (also obtainable from independent observations), the metallicity of the inflowing gas $Z_0$ and the effective yield.  We find our model can very well characterize the overall feature of the observed metallicity profiles.  In particular, the model naturally predicts the observed flattening of the metallicity gradients in the outer regions of the disk. The flattening metallicity indicates the substantial metallicity of inflowing gas at the edge of gas disk. As a test of the model, we find that the $h_{\rm R}$ of $\Sigma_{\rm SFR}$ that are returned by the fits agree quite well with the ones obtained from independent observations. With the assumption of the true yield, our model could also provide a way to constrain the mass-loading factor of galaxies by comparison of this with the effective yield returned by the fits. 


There are also some clear limitations for our model.  First, it is an idealized model, applicable for regular disk galaxies with co-rotating inflowing gas and exponential $\Sigma_{\rm SFR}$.   Galaxies with significant ex-planar inflow or strongly deviating from exponential $\Sigma_{\rm SFR}$ will for sure not be covered. Second, one should be careful when applying our model to high-redshift galaxies, because  1) more frequent mergers are expected at high redshift with respect to local universe \citep{Genel-10, Rodriguez-Gomez-15}, 2) the gas of high-redshift SF galaxies are more turbulent and therefore may not have such well-defined disk structure \citep[e.g.][]{Wisnioski-15}. 
Third, our model does not include the effects of small- scale structures in the gas disk or galaxies, such as  spiral arms and the existence of the bar.  Including them would surely introduce variations in the radial profile of metallicity.   
It is clear that our model cannot explain the positive radial gradients in metallicity seen in a few SF galaxies, suggesting that more complex physical processes are occurring in these galaxies (for instance, bar-induced gas inflows).  
Finally, we have assumed the mass-loading factor is constant across the galactic disk, which may not be true in real case.  A radial-dependent mass-loading factor would surely modify the prediction of metallicity profile. 
We remind the readers to keep all these in mind when applying our model to the observed metallicity profiles. 

\section{Summary and Conclusion} \label{sec:7}

The accretion of gas onto disk galaxies is critical in the process of chemical enrichment.  Recent simulations suggest that co-planar radial gas inflow dominates the gas accretion and sustains the star formation in the disk and associated outflows from it \citep[e.g.][]{Peroux-20, Trapp-21}.  Observationally, there is indirect evidence in support of this scenario in the mapping of (presumed) inflowing gas with cold/warm absorbers \citep[e.g.][]{Bielby-17, Schroetter-19}.  This general scenario indicates that a typical galactic gas disk can be treated as a ``modified accretion disk".  In Paper I, we explored the possible mechanisms for the radial gas inflow, and found that MRI-induced viscosity was likely the key to the viscous inflow of cold gas in the disks and the establishment and maintenance of the observed exponential form of SF disks.   

In this work, we have looked at the metal enrichment profiles that would be expected in such a model. We adopt the usual instantaneous metal-enrichment and the instantaneous mixing approximation. 
Compared to previous works,  this scenario of gas accretion simplifies the metal enrichment model by avoiding introducing arbitrary ex-planar gas inflows.  
Instead, we combine the ex-planar inflow with ex-planar outflow into an effective outflow \citep{Fraternali-15}, which is assumed to be the instantaneous $\Sigma_{\rm SFR}$ scaled with an ``effective" mass-loading factor, i.e. $\lambda\Sigma_{\rm SFR}$.   It is assumed that the metallicity of this ``effective" outflow is the same as that of the gas within the disk at that radius.

Based on the continuity of the mass of gas and metals on the gas disk, we then obtain two basic continuity equations (see Equations \ref{eq:1} and \ref{eq:2}) that link the inflow rate, the star-formation rate $\Sigma_{\rm SFR}(r)$ and the gas-phase metallicity. 
In a steady state, these can be combined to yield an analytic expression (Equation \ref{eq:xx2}) for the radial dependence of the gas-phase metallicity $Z_{\rm gas}(r)$.

To further reduce the number of free parameters in the model and allow analytic solutions, we then impose a precisely exponential profile for $\Sigma_{\rm SFR}(r)$, as typically seen in observed galaxies \citep[e.g.][Paper I]{Casasola-17, Wang-19}. 

The main results of the model are as follows. 
\begin{itemize}

\item  We are able to give the analytic solution of the steady-state gas-phase metallicity in a very simple form (see Equation \ref{eq:9}).  This analytic solution includes three key free parameters, the scalelength $h_{\rm R}$ of the star-forming disk $\Sigma_{\rm SFR}(r)$, the metallicity of the inflowing gas $Z_0$ and the effective yield $y_{\rm eff}$.  The radial gradients of gas-phase metallicity are determined by $h_{\rm R}$ and $Z_0/y_{\rm eff}$.   We have verified that this analytic solution at equilibrium is also a good approximation for galaxies that are undergoing significant and reasonable evolution of SFR and size.  

\item We find that a central cusp in the gas-phase metallicity is likely to be present if the central sink is small ($\eta\sim$0.001). This arises because essentially all of the inflowing gas must be used up before it reaches the center.  However, even in this case, the cusp can be efficiently smoothed out by diffusion effects, especially for galaxies of smaller size.
 
\item  According to our simple model, the negative gradient of gas-phase metallicity is a natural consequence of the radial inflow of cold gas.  The inflowing gas is continuously enriched by in-situ star formation on the disk as it slowly spirals in towards the disk center. 

\item Interestingly, the surface density of gas $\Sigma_{\rm gas}(r)$, or equivalently the star formation efficiency, does not enter into the solution for the enrichment: only the star-formation profile $\Sigma_{\rm SFR}(r)$ matters. This is because the $\Sigma_{\rm gas}(r)$ also determines the radial inflow velocity and thus the duration of enrichment at a given radius.

\item  If the co-planar inflow is significant, as assumed in our model, then it is more revealing to measure the gradients of gas-phase metallicity in terms of a normalized radius relative to the scale length of the star-forming disk (i.e. dex/$h_{\rm R}$) rather than in an absolute radius (i.e. dex/kpc).  

\item We fit the observed metallicity profiles of six nearby galaxies in the literature with well-measured metallicity profiles extending to very large radii.  Our model can well characterize the overall features of these observed metallicity profiles.   The observed profiles usually show a floor in metallicity in the outer regions of the disks, which is a natural prediction of our model corresponding to the metallicity of inflowing gas.
Further, we find the returned $h_{\rm R}$ of $\Sigma_{\rm SFR}$ agree well with those obtained from independent observations, which strongly supports our model. 

\end{itemize}

Our idea of treating galaxy disks as viscous ``modified accretion disks" was originally inspired by hydrodynamical simulations and by observations of the distribution of gas around galaxies \citep[e.g.][]{Kacprzak-12, Bielby-17, Schroetter-19, Mitchell-20, DeFelippis-20, Peroux-20, Trapp-21}. 
In Paper I, we showed that such disks will establish an exponential profile if the viscous stress within the disk is more or less constant with radius, and also argued that magnetic stresses from magneto-rotational instability are the likely origin of this viscosity.

In the present work, we have found that the same ``modified accretion disk model" is also consistent with the observed radial profiles of the gas-phase metallicity. The radial gradient of metallicity is a natural consequence of the radial gas inflow in the model.  Our findings suggest that the ``modified accretion disk" provides a useful framework to consider the formation and evolution of galactic star-forming disks.

\bibliography{rewritebib.bib}
\end{document}